\documentclass[conference]{IEEEtran}
\IEEEoverridecommandlockouts

\usepackage{cite}
\usepackage{amsmath,amssymb,amsfonts}
\usepackage{algorithmic}
\usepackage{graphicx}
\usepackage{textcomp}
\usepackage{xcolor}
\usepackage{colortbl}

\usepackage{amsmath,amsfonts}
\usepackage{algorithmic}
\usepackage{array}
\usepackage{textcomp}
\usepackage{stfloats}
\usepackage{url}
\usepackage{verbatim}
\usepackage{graphicx}
\usepackage{cite}
\usepackage[T1]{fontenc}
\usepackage{tabularx}
\usepackage{graphicx}
\usepackage[cmintegrals]{newtxmath}
\usepackage{algorithmic}
\usepackage{caption}
\usepackage{cite}
\usepackage{setspace} 

\usepackage{subcaption}
\usepackage{xcolor}
\usepackage[para]{footmisc}
\usepackage{tabularx,array}
\usepackage{multirow}
\usepackage{booktabs}
\usepackage[linesnumbered,ruled,vlined]{algorithm2e}
\SetKwInput{KwInput}{Input}                
\SetKwInput{KwOutput}{Output}  
\SetKwInput{KwFunction}{Function}  
\SetKwInput{KwReturn}{Return}  
\usepackage{textcomp}
\usepackage{linguex}
\usepackage[linguistics,edges]{forest}

\usepackage[flushleft]{threeparttable} 
\usepackage{booktabs,caption}

\usepackage{metalogo}

\usepackage{url}
\usepackage{csquotes} 
\usepackage{arydshln} 
\usepackage{color, soul}
\usepackage{amssymb}
\usepackage{makecell}
\usepackage{etoolbox}
\usepackage{threeparttable}
\usepackage{pifont}

\usepackage{float}
\usepackage[normalem]{ulem}
\usepackage{mathtools}

\makeatletter
\IEEEtriggercmd{\reset@font\normalfont\fontsize{7pt}{8.40pt}\selectfont}
\makeatother
\IEEEtriggeratref{1}

\usepackage{comment}

\usepackage{balance}

\begin{document}

\title{{\huge DOSM: Demand-Prediction based Online Service Management for Vehicular Edge Computing Networks }}

\author{\IEEEauthorblockN{Anum~Talpur~and~Mohan~Gurusamy}
	\IEEEauthorblockA{\textit{Department of Electrical and Computer Engineering,} \\
		\textit{National University of Singapore, Singapore}\\
		Email: anum.talpur@u.nus.edu, gmohan@nus.edu.sg}
}

\maketitle

\begin{abstract}
In this work, we investigate an online service management problem in vehicular edge computing networks. To satisfy the varying service demands of mobile vehicles, a service management framework is required to make decisions on the service lifecycle to maintain good network performance. We describe the service lifecycle consists of creating an instance of a given service (\textit{scale-out}), moving an instance to a different edge node (\textit{migration}), and/or termination of an underutilized instance (\textit{scale-in}). In this paper, we propose an efficient online algorithm to perform service management in each time slot, where performance quality in the current time slot, the service demand in future time slots, and the minimal observed delay by vehicles and the minimal migration delay are considered while making the decisions on service lifecycle. Here, the future service demand is computed from a gated recurrent unit (GRU)-based prediction model, and the network performance quality is estimated using a deep reinforcement learning (DRL) model which has the ability to interact with the vehicular environment in real-time. The choice of optimal edge location to deploy a service instance at different times is based on our proposed optimization formulations. Simulation experiments using real-world vehicle trajectories are carried out to evaluate the performance of our proposed demand-prediction based online service management (DOSM) framework against different state-of-the-art solutions using several performance metrics.  

\end{abstract}


\section{Introduction}
Vehicular edge computing networks are becoming popular in recent years \cite{EdgeIOV}. Vehicular applications are categorized under the 5G ultra-reliable and low latency communication (URLLC) slice which require vehicles to be connected in real-time with the infrastructure to get assistance in different low-latency driving tasks. Edge computing is known to assure faster service availability by bringing computing resources in close proximity to vehicles. Nonetheless, providing a seamless connection to highly mobile vehicles while maintaining low-latency service availability requires efficient service management which dynamically relocates the services in real-time as vehicles move across the service region. This paper focuses on the aspects of online service management in an edge-enabled vehicular environment. \par 
Several relocation techniques have been proposed in the literature but most of the existing approaches are reactive and suitable for delay-tolerant vehicular applications and not very effective when vehicles need to perform delay-sensitive tasks. To achieve delay-efficient network performances, the research propose to maintain collaborative service relocation/migration within edge nodes \cite{EdgeCollaboration,EdgeCollaboration2}. The existing service migration techniques use different performance parameters including, delay \cite{delay}, energy consumption \cite{MigEnergy,MigEnergy2}, signalling cost \cite{MigCost} and so on. Mada et al. \cite{liveMG} propose service migration scheme across centralized cloud and edge cloud, and to cope with the varying mobility patterns, this work proposes to always reoptimize migration decision during each time slot without any prior knowledge on the need for service migration. Recent studies have considered proactive approaches as a better solution \cite{MigEnergy2} where prediction of different user/network parameters (like distance prediction \cite{distance}, probabilistic mobility prediction \cite{mobiltiy} cost prediction \cite{cost}, and so on) is carried out before making migration decisions. Some of this existing work is in the context of cloud computing which cannot be applied for edge-enabled networks \cite{cost,cost2}. \par 
With that, most of the current works are reactive with different performance requirements, and existing studies on proactive approaches either use probabilistic structures or the choice of prediction parameter is different from the one considered in our framework. Not only this, in our work other than migration, two new actions i.e. deletion/termination (scale-in) or instantiation of an instance (scale-out) on the same egde node are considered while performing service management. Here, scale-in helps to free up underutilized edge resources and the scale-out tries to enhance the network performance by instantiating a brand-new service instance on the same edge before deciding to transfer an existing instance to avoid the disruption. Moreover, our design aims to work in an online manner for seamless provisioning of services to the vehicles. \par 
In this paper, we propose demand-prediction based online service management (DOSM) framework that works in conjunction with varying service demands with the aim of reducing service delay and migration delay while performing service lifecycle decisions (i.e. migrate, scale-in or scale-out). The key aspects of our design framework are: 1)\textit{ demand prediction}: We use a light-weight and time-efficient gated recurrent unit (GRU)-based neural network to predict the service demand and estimate the future utilization of service resources. 2) \textit{decision making network}: This takes the estimated future load and current network performance to make a decision on the service lifecycle. 3) \textit{Deep reinforcement learning (DRL) agent}: Here, we incorporate DRL into our framework to take account of real-time network performance and possible impacts of vehicle mobility. We exploit the actor-critic framework where the actor uses optimization mathematics to calculate optimal edge location for migration/deletion/instantiation of service instance and critic to estimate the network performance quality in terms of the observed service delay. To demonstrate the effectiveness of our proposed framework, we carry out simulation experiments with real-world vehicle trajectories and realistic network constraints. 
\section{System Model}
\label{Sec:SystemModelAndServiceLifecycle}
In this work, we consider a vehicular edge computing network with 5G coverage. It consists of a set $E$ of edge nodes, each co-located with evolved nodeB (eNB), and a set $S$ of services that need to be placed/deployed at edge nodes. Each edge node $e\epsilon E$ has its memory storage $\phi_e^{max}$ and computation capacities $f_e (Hz)$ to describe CPU frequency that edge node can allocate, and each instance of service $s\epsilon S$ has its input data size $D_i (bits)$, the computation intensity $C_s (cycles/bit)$, the storage requirement $\phi_s (bytes)$ and maximum allowed delay $T_s^{th}(sec)$. We have a set $V$ of vehicles that move frequently and request low-latency services to accomplish different vehicular tasks. The total number of vehicles requesting for services $s$ from edge $e$ is indicated as service demand $V_e^s$. Using 5G network connections, the vehicles associate with the nearest edge node and get access to the requested service. \par 
\begin{figure}[htbp]
	\begin{center}
		\includegraphics[width=3.2in,height=1.7in]{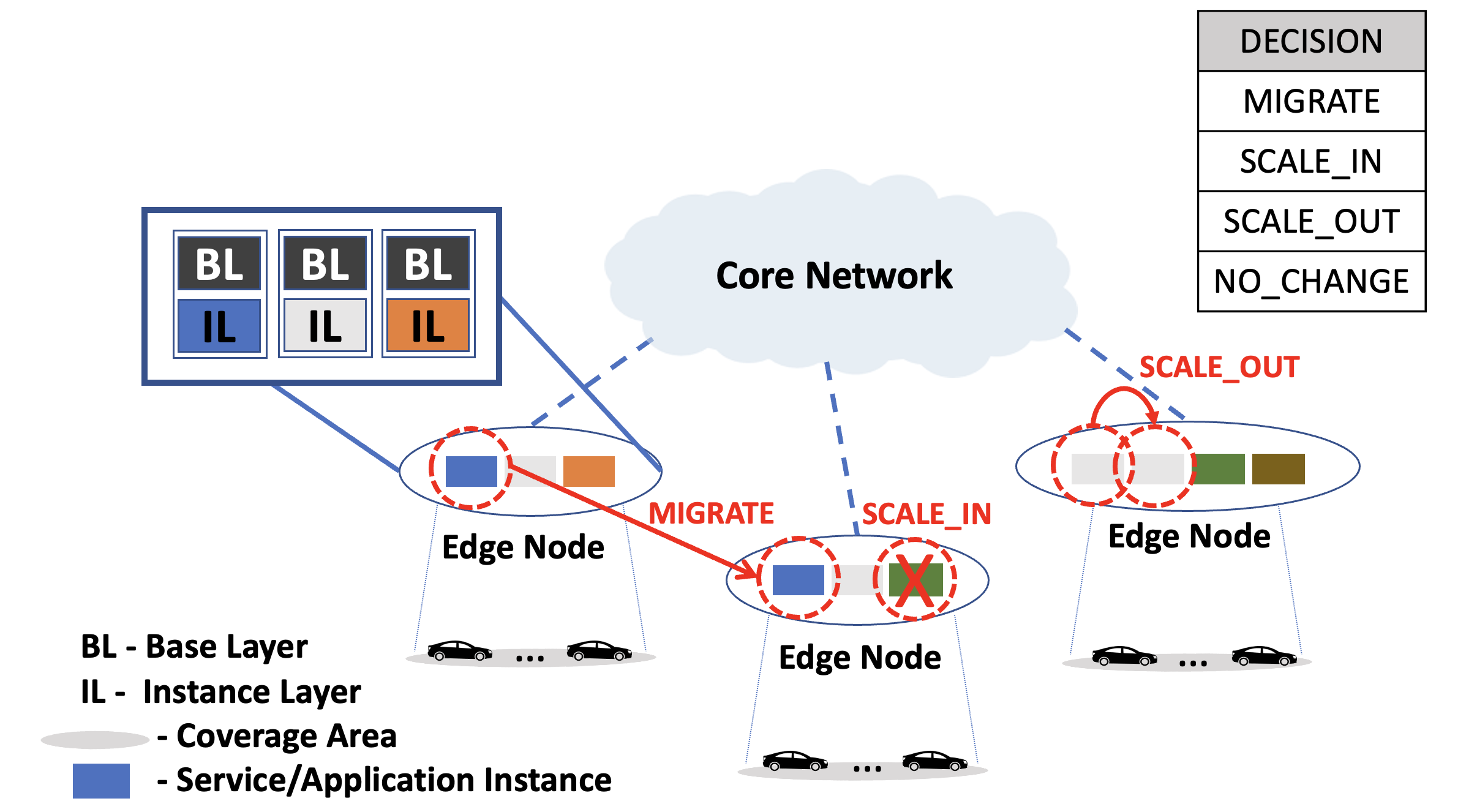}
		\caption{The System Model}
		\label{fig:Archi}
	\end{center}
\end{figure}
A service/application is a facility like collision risk, intersection safety, platoon, and so on (defined by European Telecommunications Standards Institute (ETSI)), and each service instance is a virtual machine (VM) image. To maintain delay-efficient service availability, the placement and relocation of service instances must be handled as per the vehicle's movement in an online manner. Therefore, each service has a lifecycle controlled by our proposed DOSM framework which resides over the edge management platform, according to ETSI architecture framework \cite{etsi2022multi}. Our DOSM framework benefits from the use of two-layer architecture for running services/applications \cite{liveMigration,liveMigrationContainer}, as shown in Fig. \ref{fig:Archi}. It consists of the base layer (BL) that includes the operating system, the system kernel, and so on, and the instance layer (IL) containing the application-specific data and its running state. The layered VM images in our DOSM framework enable fast relocation of services where only IL is relocated/placed and the transfer of common BL is avoided for which we assume all edge nodes share a copy of it. \par 
Without loss of generality, we assume a time-slotted system structure corresponding to $t=1,2,....\mathcal{T}$ where our DOSM framework performs periodic checks to monitor the current availability of services to the vehicles and ensures that placement of services is adaptive to traffic variations. One slot lasts for $\tau$ seconds. At each time unit $t$, the network collects the requests generated for each service type within the coverage of all edge nodes and determines the network performance for the current time unit. Next, by intelligently making the prediction of traffic for the next time unit, the service management framework will decide on one of the below-given decisions prior to the traffic starting to avail that service. The prediction of traffic is important here to keep up with the IoV environment where the services belong to the URLLC slice and the sub-second reaction times are needed. Overall, the service lifecycle decision set consists of the following:
\begin{itemize}
	\item MIGRATE: This involves moving a service instance layer from one edge node to another edge node.
	\item SCALE-IN: This involves deletion of a service instance layer from an edge node where it is underutilized
	\item SCALE-OUT: This involves the instantiation of another instance of the same service type on the same edge node where the demand is high.
	\item NO-CHANGE: This indicates no change in service deployment is needed within the network.
\end{itemize}

\section{Demand-Prediction based Online Service Management (DOSM) Framework}
\begin{figure}[htbp]
	\begin{center}
		\includegraphics[width=3.5in,height=1.95in]{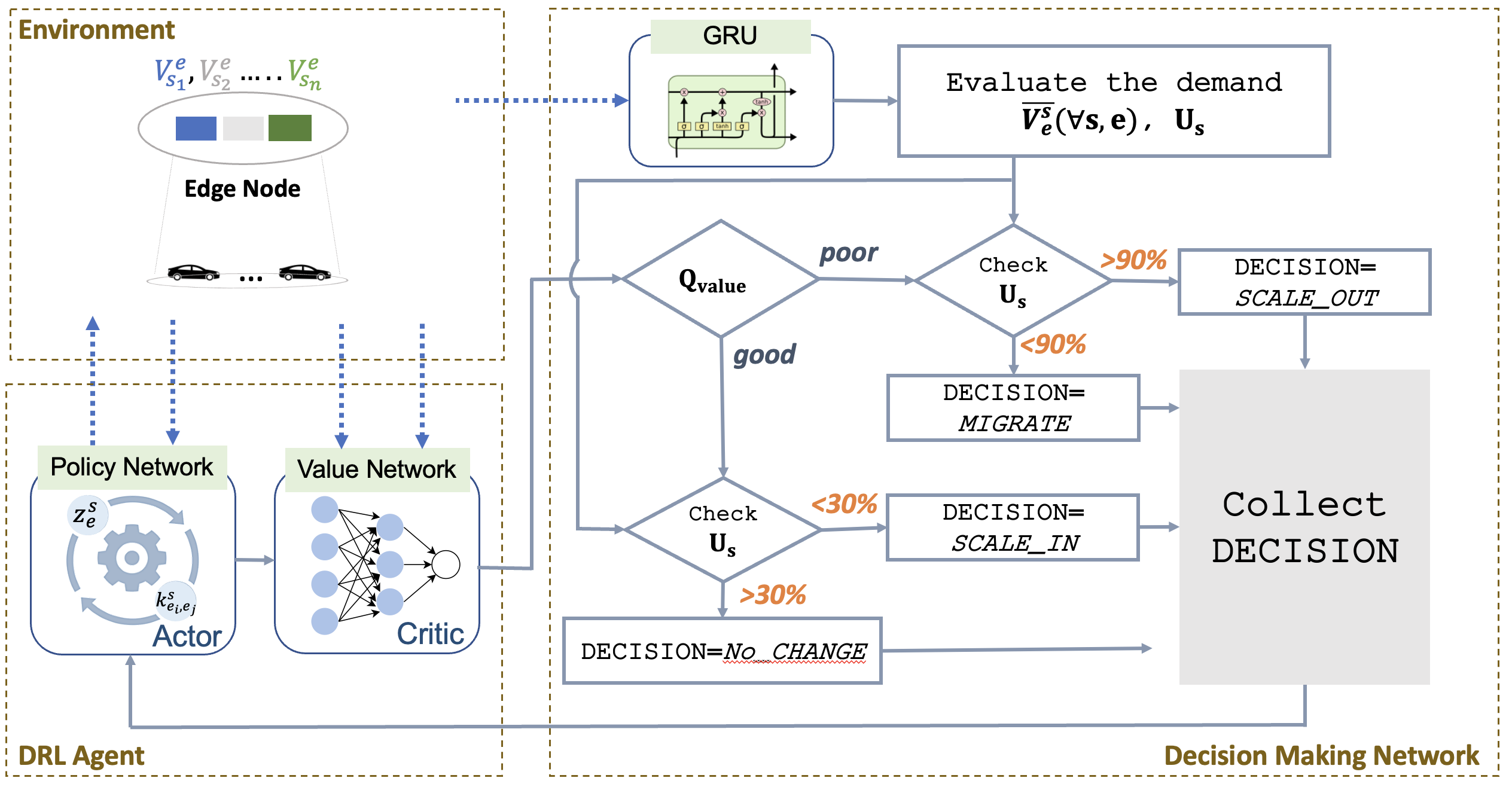}
		\caption{The Architecture of DOSM Framework}
		\label{fig:framework}
	\end{center}
\end{figure}
We depict the architecture of our proposed DOSM framework in Fig. \ref{fig:framework}. The key concept of the DOSM framework is to work in an online manner. For every $\tau$ seconds, the network performs checks and collects the service requests from the environment. Our environment consists of a real traffic of taxis in the city of San Francisco \cite{sanfrancisco} where vehicles are moving and associating with the nearest edge node and generating a continuous set of requests which require real-time analysis and has different requirements on service quality in terms of delay. Our DOSM framework has two major components- decision-making network and a DRL agent.\par 

\subsection{Decision Making Network}
\label{Sec:Framework}
This part of the network makes a predictive decision (i.e. four decisions discussed in Section \ref{Sec:SystemModelAndServiceLifecycle}) that needs to be performed over the service's lifecycle. We use GRU, a variant of recurrent neural network, because of its ability to store states over time and learn temporal information to predict future demand data. The choice of GRU is also because of its computational efficiency and low memory overhead which makes it lightweight and time-efficient in predicting data. The key idea of the predictive decision is to look into \textit{service demand} (i.e. the number of vehicles requesting service $s$ from different edge nodes) ahead of the time frame and then take the decision accordingly. The predictive decision is repeated every time frame $\mathbb{T}$ where each time frame consists of $n\times \tau$ time slots. Each time frame corresponds to a batch of service requests (input values) received in it. We pre-process the collected service request data using Z-score standardization, to scale it into a definite range and prevent training data from diverging. Next, for a given batch size, a set of future service demands $\overline{V_e^s}$ is predicted which will help to compute the percent utilization $\mathbb{U}_{s}$ of service resources. The $\mathbb{U}_{s}$ is calculated as,
\begin{equation} 
	\mathbb{U}_{s} (\%)=\frac{\overline{V_e^s}}{I_s\mathbb{C}}\times100,  \forall s\in S, e\in E 
	\label{EQ:Ucomp}
\end{equation}
Here, $I_s$ is the total number of instances deployed for service $s$, and $\mathbb{C}$ is the total number of vehicles each instance can handle or provide parallel connections. In this paper, we use a batch size of 150 which predicts a set of 15 (i.e. 10\% of 150) values for the next 15 time slots (i.e. $\mathbb{T}=15\times\tau$) with good accuracy. After each time frame $\mathbb{T}$, we roll it by one frame, update the batch data by including a new set of service requests received and make a new set of predictions. This process runs repeatedly in a continual manner intending to extend the acquired stream of knowledge. \par 
Once we have a set of predictions, the entire decision process becomes simple which is summarized in Fig. \ref{fig:framework}. The network runs continuously and performs data acquisition from an environment to sample a batch of input data. The collected data is pre-processed and used as an input of GRU model to predict a set of future service demands. The next step is to make use of a trained critic network (from the DRL agent, explained in Section \ref{Sec:DRLAgent}) and calculate the network performance in terms of $Q_{value}$. In case of poor network quality and high service utilization (i.e. > $90\%$), the \textbf{instantiation} of a new service instance is needed (with $\mathbf{DECISION}$=SCALE\_OUT). On the contrary, if service utilization is not very high (i.e. < $90\%$), the service \textbf{migration} is needed over the new optimal edge location (with $\mathbf{DECISION}$=MIGRATE) to enhance network performance. On the other hand, if the network quality is good and service utilization is low (i.e. < $30\%$), the underutilized instance can be \textbf{deleted} if more than 1 instance exists for that service type (with $\mathbf{DECISION}$=SCALE\_IN). If the service utilization is more than 30\% with good $Q_{value}$ then the network will continue with the same deployments, and no change is needed. Note that the decision is further sent to the policy network of the DRL agent to decide on an optimal edge node location where migration/deletion/instantiation of a service instance is performed.  

\vspace{-3mm}
\subsection{DRL Agent}
\label{Sec:DRLAgent}
Driven by the time-varying nature of vehicular networks, it is imperative that the real-time environment be taken into account while performing service management. We choose to integrate DRL in our framework to address the changing network conditions in terms of performance quality and use it to make effective service management decisions through time. The DRL consists of an agent and environment where the agent interacts with the environment to observe state $\delta_t$, take an action $\mathfrak{a}$, and finally, in response, a feedback $\mathcal{F}$ is received which helps to make future decisions such that good network performance is maintained. Here, network state $\delta_t$ at time unit $t$ corresponds to the vehicle identification, requested service type, and the vehicle location. We exploit the actor-critic architecture of DRL which contains \textit{policy network} and \textit{value network}.\par 
\subsubsection{Policy Network}
In our design, the \textit{\textbf{policy network}} takes the output of the decision making network as input to generate an action that determines where (i.e. location of optimal edge node) to migrate/instantiate/delete a service instance based on the current network state. We use integer linear programming (ILP) formulation to determine the optimal edge location. We start with the initial placement of services and define it as $x_e^s$ that are used at the beginning of the network as zero$^{th}$ time ($t=0$) configurations. Here, $x_e^s$ is 1 if edge node $e$ hosts service $s$; otherwise, it is 0. Given a set of services and $I_s$ number of service instances for each service type with their computation and delay requirements, the $x_e^s$ finds the optimal choice of edge servers to place the service instances at the beginning of the network. The calculation of $x_e^s$ is done from our previous work in \cite{DRLD} that finds the optimal edge servers for the initial placement of services with the objective of minimizing the maximum edge resource usage and service delay. \par
Using $x_e^s$ as input, we aim to minimize the \textbf{\textit{service delay $T_s$}} and \textbf{\textit{migration delay $T_m$}} while calculating the new optimal edge locations based on the decision we have from decision making network.\par 

\textit{Definition 1 (Service Delay ($T_s$)):} Service delay is the delay observed by vehicles while accessing service $s$ from edge node $e$ and it includes propagation delay, transmission delay, and computation delay (i.e. $T_s = T_{prop} + T_{trans} + T_{comp}$). As a rule of thumb, the propagation delay $T_{prop}$ is the ratio of the distance between vehicles and edge nodes where the instance of service $s$ is deployed, and the propagation speed over the medium. Whereas, the transmission delay $T_{trans}$ is the ratio between the input data size (of service request packet) and the transmission rate.
\begin{equation}
	T_{trans} = \frac{D_i}{R_x(v,e)}
	\label{eqTtrans1}
\end{equation}
The transmission rate between vehicle $v$ and edge node $e$ where service $s$ is deployed is, 
\begin{equation}
	R_x(v,e) = W \log_2 \left( 1 + \frac{dist(v,e)^{-2}P}{N_o} \right)
	\label{eqTtrans2}
\end{equation}
Here, $W$, $P$, and $N_o$ are the channel bandwidth, transmission power, and noise power, respectively. The computation delay $T_{comp}$ over edge node $e$ to process the service $s$ request is: \par 
\begin{equation}
	T_{comp} = \frac{D_iC_{s}}{f_e}
	\label{eqComp}
\end{equation}
\hspace{3mm}\textit{Definition 2 (Migration Delay ($T_m$)):} In our model, it is the transmission delay observed while migrating an instance of service from one edge node to another edge node. 
\begin{equation}
	T_{m} = \frac{\phi_m}{R_x(e_i,e_j)}
	\label{eqTm}
\end{equation}
Here, $\phi_m$ is the total size of VM data that need to be migrated which is equal to the sum of instance layer size ($\phi_s$) and user context ($\phi_v\times V_e^s$)($\phi_v$ is the storage requirement of each user) and $R_x(e_i,e_j)$ is the capacity of a typical wired WAN (wide area network) connectivity between edge nodes. 

Finally, based on the possible decision (discussed in Section \ref{Sec:SystemModelAndServiceLifecycle}) that can be sent to the actor of the policy network, the optimization formulation is divided into three solvable sub-problems. \\ \\
\textbf{MIGRATE Decision}: In this problem, an instance of service $s$ needs to be migrated to another edge node with the objective of minimal service delay and minimal migration delay. We use binary variable $z_e^s$ to define the new optimal edge location and binary variable $k_{e_i,e}^s$ to define the optimal migration link between edge node $e_i$ and edge node $e$. We mathematically formulate our objective function as:
\begin{equation}
	\underset{e_i,e}{minimize} \hspace*{3mm} T_s z_e^s +  T_m k_{e_i,e}^s 
	\label{eqObj1}
\end{equation}
\hspace*{10mm} subject to,
\begin{equation}
	z_e^s T_s \le T_s^{th},  \forall e\in E 
	\label{EQ:DelayConstraint}
\end{equation}
\begin{equation}
	z_e^s(\phi_e + \phi_m) \le \phi_e^{max}, \forall e\in E 
	\label{EQ:CompConstraint}
\end{equation}
\begin{equation}
	k_{e_i,e}^s T_m \le \tau, \forall e_i,e\in E 
	\label{EQ:migrationDelay}
\end{equation}
\begin{equation}
	z_{e}^s + k_{e_i,e}^s  = 2,  \forall e_i,e\in E
	\label{EQ:PlaceConstraintM2}
\end{equation}
\begin{equation} 
	\sum_{e\in E} k_{e_i,e}^s = 1, \forall e_i\in E 
	\label{EQ:PlaceConstraint}
\end{equation}
\begin{equation}
	\sum_{e\in E} |z_e^s - x_e^s | \le 2
	\label{EQ:PlaceConstraintM}
\end{equation}
\begin{equation} 
	\sum_{e\in E}z_e^s = I_s
	\label{EQ:IsConstraint}
\end{equation}
\begin{equation}
	k_{e_i,e}^s, z_e^s \in \{0,1\};  \   s\in S, \forall e,e_i\in E 
	\label{EQ:binaryzk}
\end{equation} 
Constraint (\ref{EQ:DelayConstraint}) ensures that the total service delay experienced by vehicles along different edge nodes is less than the given threshold. Constraint (\ref{EQ:CompConstraint}) limits the storage load over the edge node by the maximum storage capacity of that edge node while migrating new instance on it. Here, $\phi_e$ represents the pre-occupied resources of an edge node. Constraint (\ref{EQ:migrationDelay}) limits the migration delay to be less than one time slot. Constraint (\ref{EQ:PlaceConstraintM2}) ensures that an instance is migrated to edge node $e_j$ only if a low migration delay link exists to that edge node. Constraint (\ref{EQ:PlaceConstraint}) guarantees each edge node has a unique low delay link to another edge node. Constraint (\ref{EQ:PlaceConstraintM}) ensures that an instance is migrated from one edge node to another edge node only if placement decision of $x_e^s$ (old placement) is different from $z_e^s$ (new placement) for service $s$. Constraint (\ref{EQ:IsConstraint}) guarantees that total $I_s$ number of instances must be placed along different edge nodes. Finally, (\ref{EQ:binaryzk}) defines the $z_e^s$ and $k_{e_i,e}^s$ as a  binary integer decision variable.\\ \\
\textbf{SCALE\_OUT Decision}: In this problem, the instantiation of a new instance of same service type takes place at one of the optimal edge locations where instance of service $s$ is already placed. The objective is to minimize the service delay while choosing the optimal edge location, and formulated as,
\begin{equation}
	\underset{e}{minimize} \hspace*{3mm} T_s z_e^s 
	\label{eqObj2}
\end{equation}
\hspace*{10mm} subject to,  (\ref{EQ:DelayConstraint}), (\ref{EQ:CompConstraint}), (\ref{EQ:binaryzk}) and
\begin{equation} 
	\sum_{e\in E}z_e^s = I_s + 1
	\label{EQ:IsConstraintSI}
\end{equation}
\begin{equation}
	z_e^s  \le 2  x_e^s, \forall e\in E 
	\label{EQ:PlaceConstraintSI}
\end{equation}
Constraint (\ref{EQ:IsConstraintSI}) ensures the placement of one additional instance compared to the existing number of instances to perform scaling out by instantiation of another instance of the service type $s$. Constraint (\ref{EQ:PlaceConstraintSI}) gaurantees the scaling must take place at the existing edge node where service $s$ is already placed as $x_e^s$.  \\ \\
\textbf{SCALE\_IN Decision}: In this problem, the deletion of one existing and underutilized instance of service type $s$ is performed. With the similar objective as in (\ref{eqObj2}), the deletion of an instance is subject to (\ref{EQ:DelayConstraint}), (\ref{EQ:CompConstraint}), (\ref{EQ:binaryzk}) and,
\begin{equation} 
	\sum_{e\in E}z_e^s = I_s - 1
	\label{EQ:IsConstraintSO}
\end{equation}
\begin{equation}
	z_e^s  \le x_e^s, \forall e\in E 
	\label{EQ:PlaceConstraintSO}
\end{equation}
Constraint (\ref{EQ:IsConstraintSO}) ensures the placement of one less instance compared to the existing $I_s$ instances. Constraint (\ref{EQ:PlaceConstraintSO}) guarantees the deletion must take place at one of the existing edge nodes where service $s$ is already deployed at $x_e^s$. 
\vspace{-3.5mm}
\subsubsection{Value Network}
The value network takes the current state and feedback as input to estimate the network's performance quality function (i.e. $Q_{value}$). Here, feedback is a response in terms of observed service delay (i.e. $\mathcal{F}=\mathbb{E}\left[T_s(t)\right]$) generated for the corresponding action $\mathfrak{a}$ of the policy network. In our design, the value network contains three fully-connected layers and updates its parameters $\theta$ to minimize the mean square loss function $\mathcal{L}_Q$ based on the feedback $\mathcal{F}$ and its corresponding request parameters. The loss function is computed as:
\begin{equation}
	\mathcal{L}_Q(\theta)=\frac{1}{\mathcal{N}}\sum_{i=1}^{\mathcal{N}}\left[(y_{t_i}-Q_{value}(\mathfrak{a};\theta))^2\right]
	\label{eq:loss}
\end{equation}
The $y_t$ is a target value which is calculated as:
\begin{equation}
	y_t =
	\begin{cases}
		\sigma(T_s^{th},\mathcal{F}) & \mathcal{F} < T_s^{th}\\
		0 & \text{else}
	\end{cases} 
	\label{eq:yt}
\end{equation}
Here, $\sigma(T_s^{th},\mathcal{F})$ is the standard deviation between the delay threshold and observed service delay. The $Q_{value}$ changes between 0 to 1, a small value implies poor performance which is updated to the decision-making network for better management of the service lifecycle. In our design, we use a threshold of 0.5 to indicate the poor ($Q_{value}$<0.5 ) and good ($Q_{value}$>0.5) network performance. The choice of 0.5 is not random but is based on our previous study performed in \cite{DRLD}.\par 
Note that due to space limitation, we skip the basics on GRU and DRL, and interested readers are referred to \cite{GRUBasics}.

\section{Performance Evaluation}
The implementation of DOSM is carried out using MATLAB software. We consider the vehicle trajectory from real-world dataset \cite{sanfrancisco} where a maximum of 500 taxis traveling along the city from which we extract the area of $15\times 15km^2$ for our experiments. The key parameters are summarized in Table \ref{tab:parameters}. We gather possible service types and related information from ETSI standards \cite{ETSI1,ETSI2}. For the value network, we use a fully-connected network with 3 hidden layers, each with 512, 256, and 64 neurons, respectively, and the output layer is a single neuron that expresses the $Q_{value}$ with the linear transfer function. The maximum number of episodes to train a network is 1500 with each episode having a maximum of 20 iterations and a batch size of 100. For the GRU model, we use 2 GRU layers (400 and 200 hidden units), 2 fully connected layers (100 neurons each), an output layer, and a regression layer. The maximum number of epochs is set to 150 with a batch size of 150. We compare our proposed \textbf{DOSM} framework with existing migration schemes \cite{comparison1,comparison2}, including never migrate (\textbf{NM}), always migrate (\textbf{AM}), and migration using DRL without demand prediction (\textbf{DRL}). Similar to our DOSM framework, we run baselines algorithms in an online manner for fair comparison in the experiments.

\begin{table}[htbp]  
	\centering  
	\caption{Parameters}    
	\scriptsize
	\tabcolsep=0.07cm
	\begin{tabular}{ |p{9em}cccc|p{4.5em}c|}    
		\hline 
		  \textbf{Service Type} & \textbf{$T_s^{th}$} & \textbf{$D_i$} & \textbf{$C_s(K)$} & \textbf{ $\phi_s$} & \textbf{Parameter} & \textbf{Value} \\  
		\hline    
		\textbf{Emergency Stop} & 0.1   & 3200  & 36    & $U(50,150)$ & \textbf{$E$} & 9 \\    
		\textbf{Collision Risk} & 0.1   & 4800  & 40    & $U(50,150)$ & \textbf{$f_e$} & 10 GHz \\    
		\textbf{Accident Report} & 0.5   & 4800  & 28    & $U(50,150)$ & \textbf{$C_e^{max}$} &2 GB \\    
		\textbf{Parking} & 0.1   & 1200  & 80    & $U(150,300)$ & \textbf{$\phi_v$} & 1 MB \\    
		\textbf{Traffic Control} & 1     & 1200  & 45    & $U(150,300)$ & \textbf{$R_x(e,e)$} & 1Gbps \\    
		\textbf{Platoon} & 0.5   & 4800  & 88    & $U(150,300)$ & \textbf{$\mathbb{C}$} &  30  \\    
		\textbf{Face Detection} & 0.5   & 3200  & 50    & $U(150,300)$ & \textbf{$P$} & 40 dBm \\
		\textbf{Intersection Safety} & 0.05  & 1800  & 42    & $U(50,150)$ & \textbf{$N_o$} &  -100dBm \\    
		- & -  & -  & -    & - &  \textbf{$W$} & 1MHz \\    
		\hline   
	 \end{tabular}%
 \label{tab:parameters}%
\end{table}%
\vspace{-4mm}
\subsection{Results}
\begin{figure}[hbt!]
	\captionsetup[subfigure]{justification=centering}
	\centering
	\begin{subfigure}{.21\textwidth}
		\centering
		\includegraphics[width=1.5in,height=1.2in]{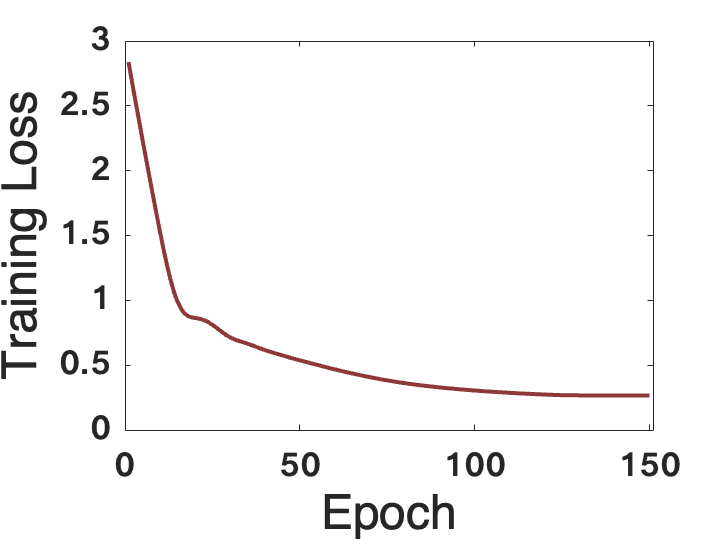}  
		\caption{Training Loss}
		\label{fig:TrainingLoss}
	\end{subfigure}
	\begin{subfigure}{.21\textwidth}
		\centering
		\includegraphics[width=1.5in,height=1.2in]{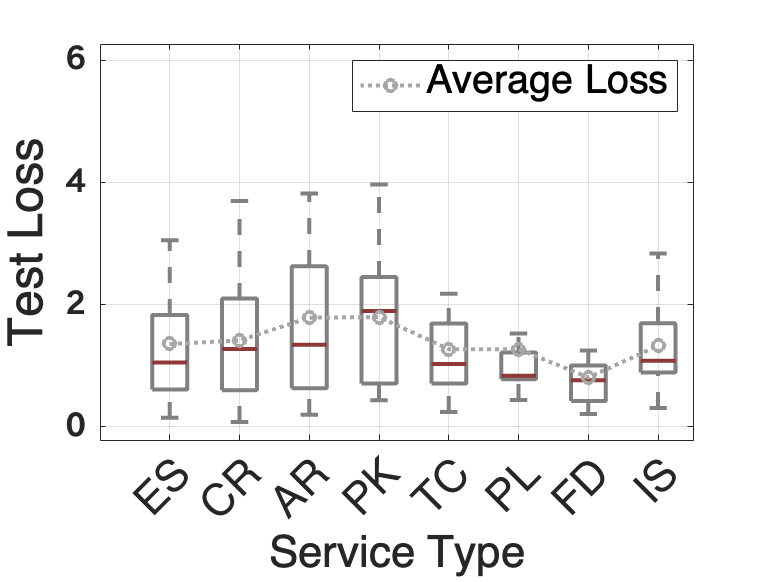}  
		\caption{Test Loss}
		\label{fig:TestLoss}
	\end{subfigure} \\
	\begin{subfigure}{.21\textwidth}
		\centering
		\includegraphics[width=1.5in,height=1.2in]{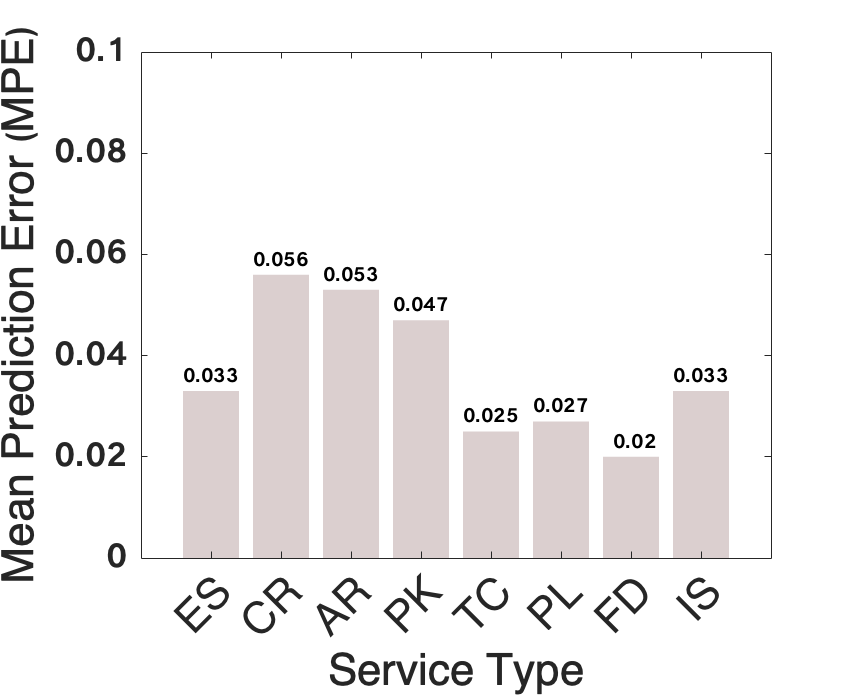}  
		\caption{MPE}
		\label{fig:MPE}
	\end{subfigure}
	\caption{GRU-based Prediction Performance}
	\label{fig:GRU}
\end{figure}

In this section, we evaluate and discuss the performance of our proposed DOSM framework using different metrics. We plot the performance of our GRU-based prediction model in Fig. \ref{fig:GRU}. Here, training loss and test loss indicate the error in the training data and test data, respectively. We use mean square error (MSE) as our loss function where the smaller the value, the better the prediction. The training loss in Fig. \ref{fig:TrainingLoss} tends to decrease with the increasing epoch whereas test loss in Fig. \ref{fig:TestLoss} for all types of services remains below the average value of 2. In addition, we plot the mean prediction error in Fig. \ref{fig:MPE} which we define as the ratio between the absolute difference of the true observed value and predicted value, and the true observed value. It can be noted that the maximum deviation of predicted value from true value ranges between 2\% (i.e. 0.02) to 5.6\% (i.e.0.056) only. The significance of low prediction error is important in our design as our proposed service management framework is based on the predicted demand. All of these performance plots for the GRU-based model indicate higher performance of our designed prediction model. \par 

\begin{figure*}[hbt!]
	\captionsetup[subfigure]{justification=centering}
	\centering
	\begin{subfigure}{.19\textwidth}
		\centering
		\includegraphics[width=1.45in,height=1.2in]{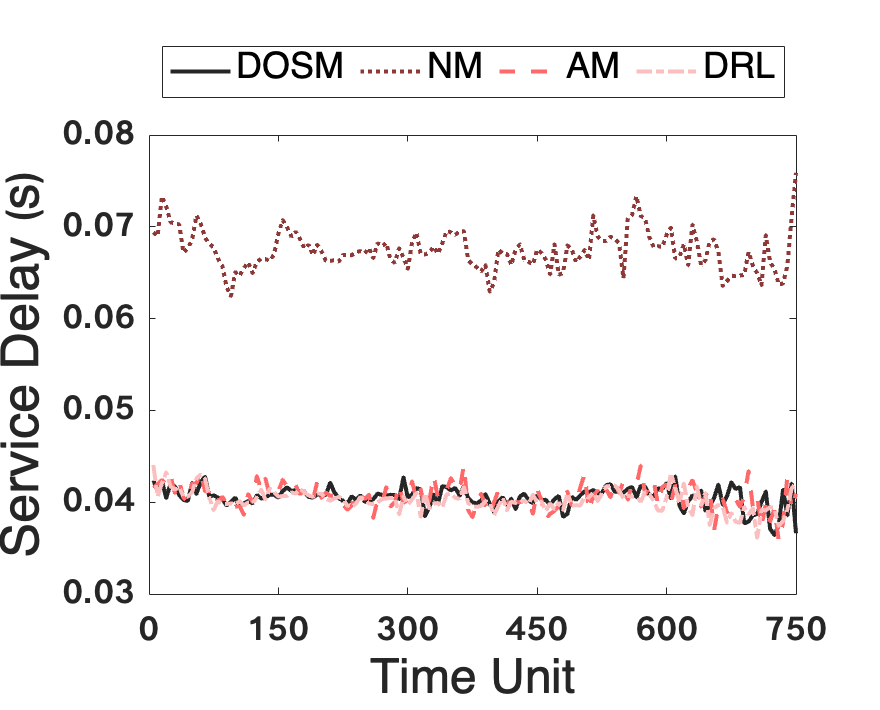}  
		\caption{Service Delay}
		\label{fig:ServiceDelay}
	\end{subfigure}
	\begin{subfigure}{.19\textwidth}
		\centering
		\includegraphics[width=1.45in,height=1.2in]{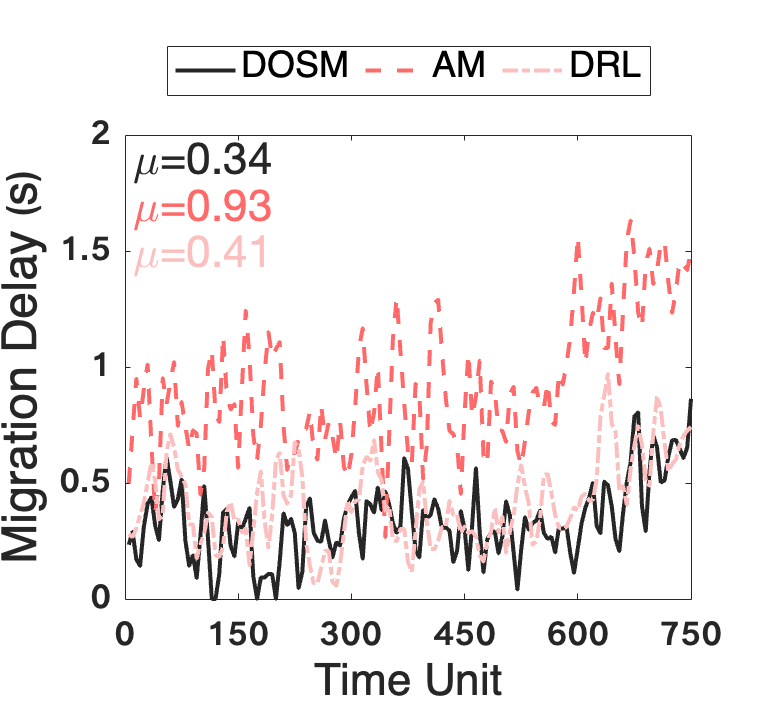}  
		\caption{Migration Delay}
		\label{fig:MigrationDelay}
	\end{subfigure} 
	\begin{subfigure}{.19\textwidth}
		\centering
		\includegraphics[width=1.45in,height=1.2in]{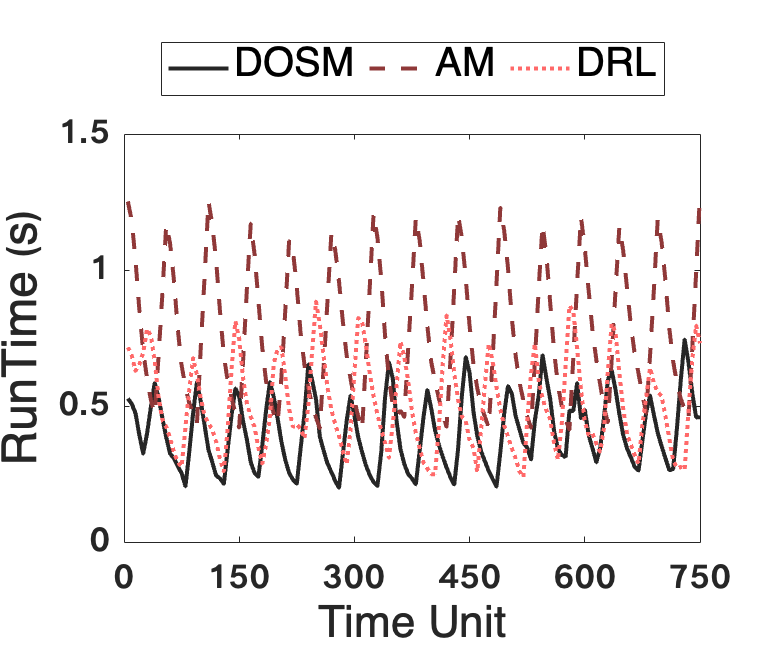}  
		\caption{RunTime}
		\label{fig:RunTime}
	\end{subfigure}
		\begin{subfigure}{.19\textwidth}
			\centering
			\includegraphics[width=1.4in,height=1.1in]{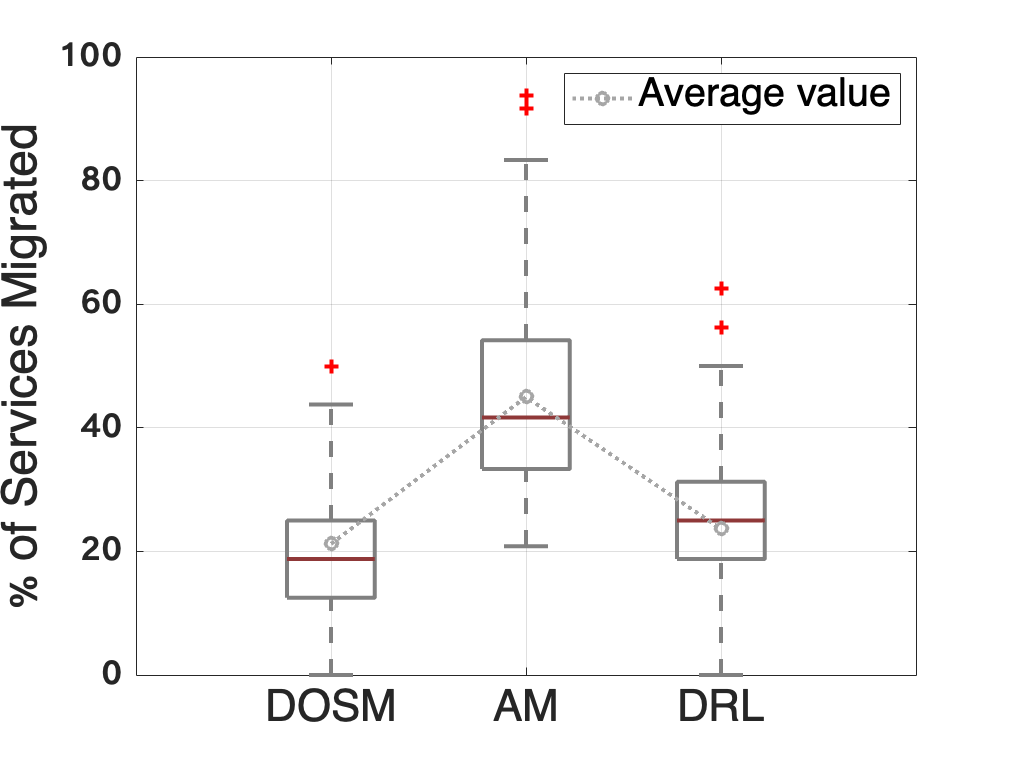}  
			\caption{Services Migrated}
			\label{fig:MigrationOverhead}
		\end{subfigure}
		\begin{subfigure}{.19\textwidth}
			\centering
			\includegraphics[width=1.35in,height=1.1in]{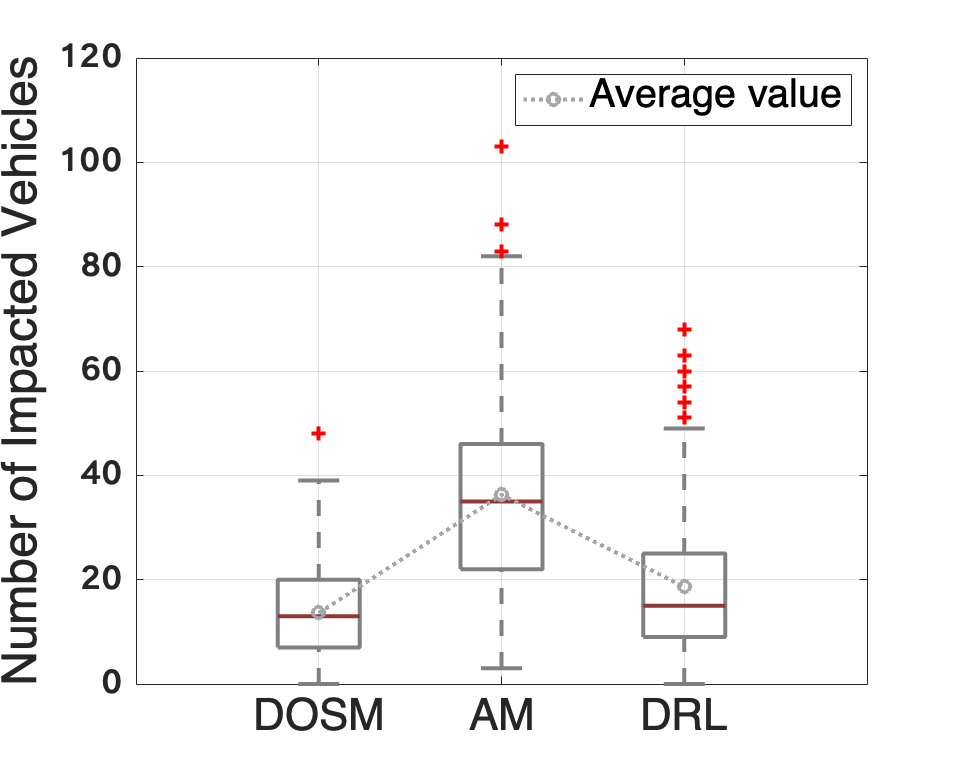}  
			\caption{Impacted Vehicles}
			\label{fig:NoOfImpactedVehicles}
		\end{subfigure} 
	\caption{Performance Results}
	\label{fig:delay}
\end{figure*}

Next, we evaluate the delay performance of our proposed DOSM framework against different baselines in Fig. \ref{fig:delay}. The average service delay during different time units is shown in Fig. \ref{fig:ServiceDelay}. As can be seen, the average delay is highest in NM where service placement is static. On the other hand, the average service delay for our DOSM framework is low and nearly similar to AM and DRL. The lower delay of AM is due to the fact that the services are always checked for migration and when needed it is migrated to the nearest edge node to follow the vehicles. The good delay performance of DRL is also reasonable because of its ability of real-time interaction with the environment to address the changing network conditions. However, with the similar service delay observed in DOSM, AM, and DRL, there are number of other performance parameters (discussed next) which do not make AM and DRL suitable compared to our proposed DOSM. \par
Fig. \ref{fig:MigrationDelay} compares the migration delay in each time slot. As can be seen, the proposed DOSM achieves the lowest delay most of the time as the number of migrations is not just dependent on the performance of the network during the current time slot but also due to the ability of DOSM to accurately predict the possible future demand before making any decision on service relocation. Also, DOSM has the lowest mean migration delay compared to AM and DRL. Fig. \ref{fig:RunTime} further compares the running time for all three algorithms during different time slots. We can observe that our proposed DOSM framework maintains the same service delay performance with lesser time complexity when compared to AM and DRL. \par

Fig. \ref{fig:MigrationOverhead} plots the possible range of percentage of services migrated during each time slot using different algorithms. Since our proposed DOSM framework triggers the migration based on the predicted demand along with the current network performance, the number of migrated services during each time slot remains comparatively low when compared with DRL and AM. On the contrary, AM and DRL consider the delay performance of the current time slot only while deciding on migration decisions, and this results in more services being migrated each time. We also plot the range of the total number of impacted vehicles due to migration during each time slot in Fig. \ref{fig:NoOfImpactedVehicles}. On an average, the number of impacted vehicles that suffer disruption due to migration is approximately 50\% less in DOSM when compared to DRL, and approximately 70\% less when compared to AM. \par
In Table \ref{tab:load}, we depict the total number of times optimization formulation is solved to find the optimal migration locations for different service types and the total number of times services are actually migrated for efficient network performance. We can observe that our proposed DOSM has a clear advantage over AM and DRL in terms of computation load and migration load. The resources at the edge are limited and keeping migration triggers as low as possible is important. From the user's perspective, fewer migrations result in lower service disruption or downtime which is significant particularly when the demand is high and network resources are congested.
\begin{table}[htbp]  
	\centering 
	 \caption{Load Performance}    
	 	\scriptsize
	 \tabcolsep=0.07cm
	 \begin{threeparttable}
	 \begin{tabular}{|l|ccc|}    
	 	\hline          
	 	& \textbf{AM} & \textbf{DRL} & \textbf{DOSM} \\    
	 	\hline    
	 	\textbf{Total Number of Optimization Runs $\mathbb{N}_{o}$} & 1200  & 798   & 612 \\   
	 	 \textbf{Total Number of Migrations $\mathbb{N}_{m}$} & 572   & 323   & 279 \\   
	 	  \textbf{Computation Load (\%)} $=(\mathbb{N}_{o}/1200\tnote{*}\ )\times100$ & 100\% & 66.50\% & 51\% \\    
	 	  \textbf{Migration Load (\%)} $=(\mathbb{N}_{m}/1200\tnote{*}\ )\times100$& 47.66\% & 26.91\% & 23.25\% \\    
	 	  \hline   
 	   \end{tabular}%
    	\begin{tablenotes}
    		\item[*] $(\mathcal{T}/\tau)\times S = (750/5)\times 8 = 1200$
    	\end{tablenotes}
	\end{threeparttable}
    \label{tab:load}%
\end{table}%

\section{Conclusion}
In this paper, we considered a vehicular edge computing network consisting of mobile vehicles that move frequently and request services from the edge network to accomplish different vehicular tasks. To maintain efficient network performance, it is important that the deployment of services must follow the vehicle's movement and be relocated to the nearest edge node in an online manner. In this regard, we studied the problem of online service management for multiple vehicular services with different design parameters and different delay requirements. We focused on minimizing service delay along with minimal migration load and minimal service interruption which may happen due to frequent migrations. We developed and integrated a proactive demand prediction model with the proposed optimization formulation to make good decisions related to service management. Finally, we carried out performance study to demonstrate the superiority of our proposed DOSM framework over the baseline methods in terms of several important performance metrics. 

\bibliographystyle{IEEEtran}
\bibliography{IEEEabrv,References} 

\balance

\end{document}